\documentclass[sigconf,nonacm]{acmart}

\AtBeginDocument{%
  }

\renewcommand\footnotetextcopyrightpermission[1]{}

\usepackage{amsmath}
\usepackage{booktabs}
\usepackage{xspace}
\usepackage{graphicx}

\newcommand{\cadd}{CADD\xspace}
\newcommand{\DACM}{DACM\xspace}

\begin{document}

\title{I'm Fine, But My Voice Isn't: Cross-Modal Affective
  Dissonance Detection for Reflective Journaling}

\author{Sumin Lee}
\affiliation{%
  \institution{Seoul National University}
  \city{Seoul}
  \country{Republic of Korea}}
\email{cirtuare@snu.ac.kr}

\begin{abstract}
Digital journaling creates an authenticity gap: users consciously
translate raw emotions into text, often sanitizing narratives even
in private writing.
We formalize this as \textbf{Cross-Modal Affective Dissonance
Detection} (\cadd), a directional three-way classification
distinguishing Masking (positive text, negative acoustics),
Coping (negative text, positive acoustics), and Congruent
utterances, grounded in Gross's process model of emotion
regulation.
We present three further contributions: (i)~\textbf{CADD-Journal}, a 1,800-sample
TTS dataset with a shared-sentence-pool design that provably
isolates acoustic signal from textual content; (ii)~\textbf{DACM},
a dual-encoder model with asymmetric cross-modal attention that
resolves a gradient degeneracy in pooled fusion, achieving
macro-F1 $0.711 \pm 0.025$---with a four-step ablation
demonstrating that asymmetric attention is the dominant driver
($+0.242$) while the DIM is effective only on cross-modal
features ($+0.033$); and (iii)~a
\textbf{domain gap quantification}: zero-shot evaluation across
three naturalistic corpora reveals a substantial gap between TTS-trained models and real
speech, and we identify two concrete requirements for future
in-the-wild corpus construction.
\textbf{ReflectJournal}, a proof-of-concept iOS application,
operationalizes the framework and provides a deployment platform
for naturalistic data collection. \footnote{Code: \url{https://github.com/cirresearch/CADD}}
\end{abstract}



\maketitle

\section{Introduction}

Expressive writing produces measurable benefits for physical and
mental health when individuals engage authentically with their
emotional experience~\cite{pennebaker1997,smyth1998}.
Yet impression management persists even in private
journaling~\cite{leary1990}: users subconsciously sanitise
narratives to maintain a positive self-image~\cite{larson1990},
and self-concealment predicts anxiety and depression.
Current journaling applications rely exclusively on text,
providing no mechanism to detect or surface this authenticity gap.

\textbf{Voice as a hidden signal.}
Prosodic features---pitch, energy, speaking rate, and
timbre---reliably convey affective states that verbal content
alone cannot capture~\cite{elayadi2011,eyben2016}.
Crucially, vocal and verbal channels frequently diverge; in
IEMOCAP~\cite{busso2008} and CMU-MOSEI~\cite{zadeh2018},
acoustic emotion regularly contradicts lexical content.

\textbf{The lexical dominance problem.}
Existing multimodal models assume cross-modal alignment and
consequently suppress the dissonance signal.
Recent work shows that audio--language models rely on lexical
semantics under acoustic--semantic conflict while
underutilising prosodic cues---a phenomenon termed
\emph{lexical dominance}~\cite{chen2026}.
Prior fusion methods progress from tensor products~\cite{zadeh2017}
to directional cross-modal transformers~\cite{tsai2019multimodal}
and contrastive alignment~\cite{yang2023,mai2023}, but typically treat cross-modal
divergence as noise to minimise rather than a signal to exploit.
\cadd reframes this failure mode as a detection target.

Specifically, our contributions are:
\begin{enumerate}
\item \textbf{CADD task}: three-way classification distinguishing
  Masking (text$+$, audio$-$), Coping (text$-$, audio$+$),
  and Congruent (aligned), grounded in Gross's process
  model~\cite{gross1998}.

\item \textbf{CADD-Journal}: a 1,800-sample TTS dataset whose
  shared-sentence-pool design ensures any F1 advantage is
  genuinely multimodal.

\item \textbf{DACM architecture}: a dual-encoder model with
  asymmetric cross-modal attention resolving a gradient
  degeneracy in pooled fusion, and a Dissonance Interaction
  Module (DIM) producing an interpretable mismatch score~$S$.

\item \textbf{Domain gap quantification}: zero-shot evaluation
  across three naturalistic corpora reveals a substantial gap
  between TTS dissonance and real speech, providing, to our
  knowledge, the first systematic quantification of this gap,
  together with two concrete requirements for future
  naturalistic corpus construction. \textbf{ReflectJournal},
  an iOS design probe, operationalizes the framework and
  provides a deployment platform for that collection.
\end{enumerate}

\section{Task Definition}
\label{sec:task}

Given $(T_i, A_i)$---a text transcript and speech
recording---\cadd requires $f_\theta: (T,A) \to \{0,1,2\}$
over $\mathcal{D}=\{(T_i, A_i, y_i)\}_{i=1}^N$.
Table~\ref{tab:taxonomy} defines the taxonomy drawing on
Gross's process model~\cite{gross1998} as a conceptual lens;
labels are interpretive anchors rather than claims that
TTS-rendered samples instantiate genuine emotion regulation.
The directional distinction is therapeutically meaningful:
Masking and Coping warrant different reflective prompts.
A text-only classifier collapses to predicting Congruent under
shared-sentence-pool conditions, validating that any F1 gain
must exploit acoustic information. Consequently, we use macro-F1 as our primary evaluation metric.

\begin{table}[htbp]
\centering\small
\caption{CADD affective taxonomy.}
\label{tab:taxonomy}
\begin{tabular}{lccc}
\toprule
Class & Condition & Regulation & Label \\
\midrule
Masking   & text$+$, audio$-$ & Suppression & 0 \\
Coping    & text$-$, audio$+$ & Reappraisal & 1 \\
Congruent & aligned          & ---         & 2 \\
\bottomrule
\end{tabular}
\end{table}

\section{DACM}
\label{sec:DACM}

\subsection{Architecture Overview}

\begin{figure}[htbp]
  \centering
  \includegraphics[width=\columnwidth]{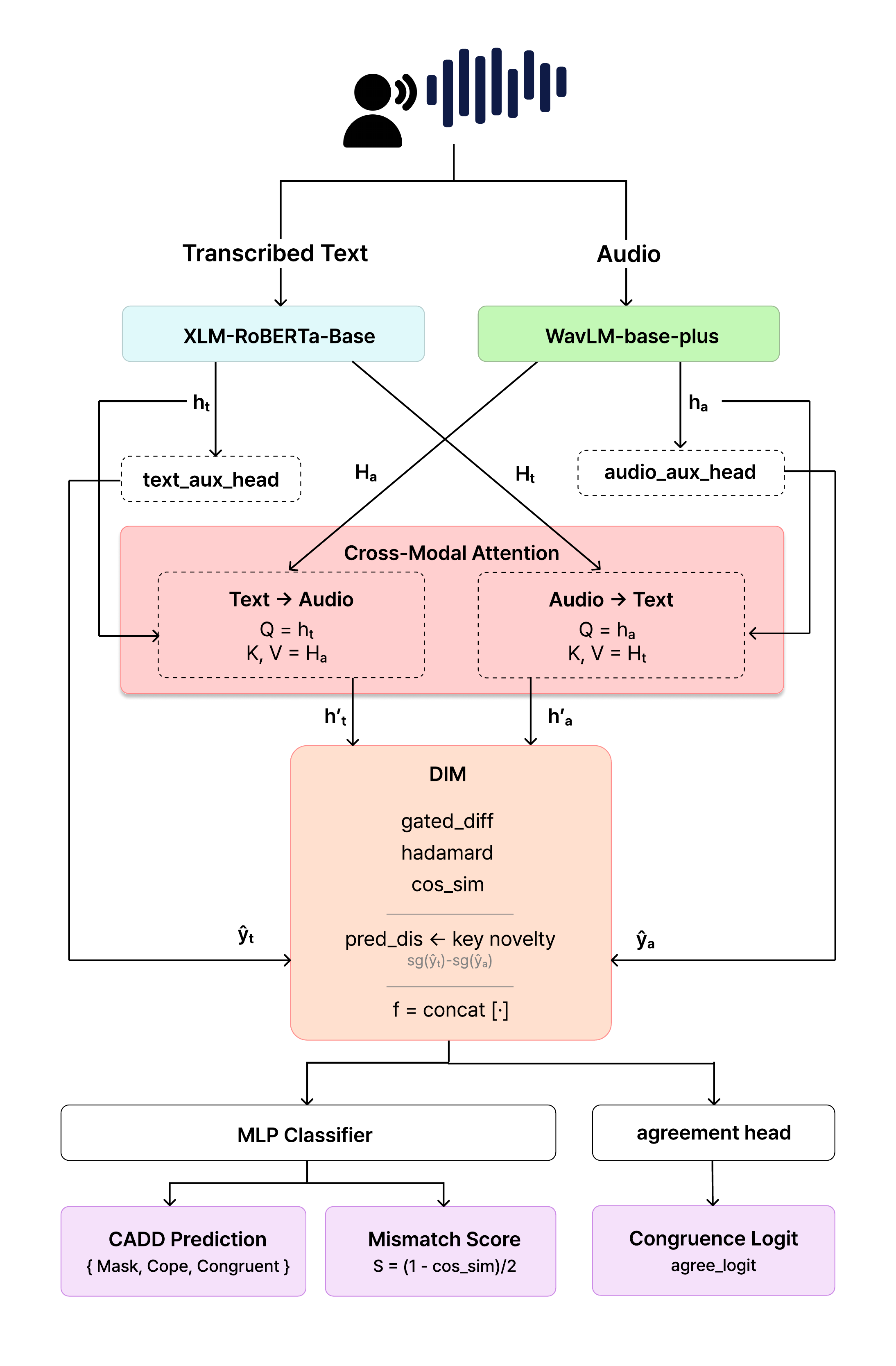}
  \caption{\DACM\ architecture. Frozen text (blue, XLM-RoBERTa)
    and audio (green, WavLM) encoders feed Asymmetric
    Cross-Modal Attention (pink), in which each modality's
    pooled vector queries the opposing modality's full sequence;
    this design resolves the scalar-softmax gradient degeneracy
    of pooled-to-pooled attention (\S\ref{sec:attn}). The
    Dissonance Interaction Module (orange) computes gated
    difference, Hadamard product, cosine similarity, and a
    stop-gradient unimodal-disagreement vector, producing the
    CADD class label and the interpretable mismatch score
    $S\in[0,1]$ (purple).}
  \Description{A vertical flowchart with two parallel encoder
  paths (text on the left, audio on the right). Each encoder
  emits a pooled vector and a full sequence. The pooled vector
  of one modality is used as a query against the full sequence
  of the other modality in a cross-modal attention block in
  the middle. The two resulting cross-attended vectors enter a
  Dissonance Interaction Module that computes four interaction
  features. These features feed an MLP classifier producing
  three CADD class logits, and a separate agreement head
  producing a binary congruence logit and the mismatch score S.}
  \label{fig:architecture}
\end{figure}

We name our model DACM (Dissonance-Aware Cross-Modal model).
Figure~\ref{fig:architecture} illustrates four sequential
stages: (1)~unimodal encoding with Weighted Layer Pooling
(WLP), (2)~pre-attention auxiliary heads,
(3)~Asymmetric Cross-Modal Attention, and
(4)~the Dissonance Interaction Module (DIM).

\subsection{Unimodal Encoders}

Text is encoded by XLM-RoBERTa-base~\cite{conneau2020},
producing pooled CLS vector $\mathbf{h}_t \in \mathbb{R}^{768}$
and full token sequence $\mathbf{H}_t$.
Audio is encoded by WavLM-base-plus~\cite{chen2022},
producing masked-mean vector $\mathbf{h}_a \in \mathbb{R}^{768}$
and frame sequence $\mathbf{H}_a$.
WLP~\cite{yang2021} computes a learnable weighted average across
all 12 hidden layers; both encoders are frozen, ensuring
identical encoder capacity across all ablation conditions.

\subsection{Asymmetric Cross-Modal Attention}
\label{sec:attn}

Standard pooled-to-pooled attention is degenerate. Specifically, the
attention score reduces to a scalar,
$\mathrm{softmax}(\cdot)\equiv 1$, so $W_Q$ and $W_K$
receive zero gradient.
We resolve this by using one modality's pooled vector as
query and the opposing modality's full sequence as keys
and values:
\begin{align*}
\mathbf{h}_{t \to a} &=
  \mathrm{MHA}(Q{=}\mathbf{h}_t,\; K{=}V{=}\mathbf{H}_a),\\
\mathbf{h}_{a \to t} &=
  \mathrm{MHA}(Q{=}\mathbf{h}_a,\; K{=}V{=}\mathbf{H}_t).
\end{align*}
The score tensor $[B \times H \times 1 \times T']$ yields a
non-trivial softmax and meaningful gradients for both
$W_Q$ and $W_K$. Unlike MulT~\cite{tsai2019multimodal}, which uses full sequence-to-sequence
cross-attention, our pooled-query design is the specific choice
that eliminates the scalar-softmax degeneracy.

\subsection{Dissonance Interaction Module}
\label{sec:dim}

Let $\mathbf{h}'_t = \mathbf{h}_{t \to a}$ and
$\mathbf{h}'_a = \mathbf{h}_{a \to t}$, both in $\mathbb{R}^{768}$.
The DIM computes four features:
\begin{align*}
\mathbf{g} &= \sigma(W_g[\mathbf{h}'_t;\mathbf{h}'_a])
             \odot(\mathbf{h}'_t - \mathbf{h}'_a), \quad
\mathbf{m} = \mathbf{h}'_t \odot \mathbf{h}'_a, \\
s          &= \cos(\mathbf{h}'_t,\, \mathbf{h}'_a), \quad
\mathbf{d} = \mathrm{sg}(\hat{y}_t) - \mathrm{sg}(\hat{y}_a),
\end{align*}
with $\mathbf{g},\mathbf{m}\in\mathbb{R}^{768}$,
$s\in\mathbb{R}$, and $\mathbf{d}\in\mathbb{R}^{3}$
(unimodal class-logit difference);
$\mathrm{sg}(\cdot)$ denotes stop-gradient, ensuring
$\mathbf{d}$ captures genuine unimodal disagreement rather than
a back-propagated artefact of $\mathcal{L}_{\mathrm{CE}}$.
The \textbf{mismatch score} $S = \tfrac{1-s}{2} \in [0,1]$
($S{\to}1$ maximally dissonant).
The concatenated feature
$\mathbf{f} = [\mathbf{g};\mathbf{m};s;\mathbf{d}]
\in \mathbb{R}^{1540}$ (i.e.\ $768+768+1+3$) feeds a two-layer MLP
(hidden~512, dropout~0.4) and a binary agreement head.

\subsection{Training Objective}

\begin{multline}
  \mathcal{L} =
  \mathcal{L}_{\mathrm{CE}}(\varepsilon{=}0.1)
  + 0.3\,\mathcal{L}_{\mathrm{margin}} \\
  + 0.2\,\mathcal{L}_{t}^{\mathrm{aux}}
  + 0.2\,\mathcal{L}_{a}^{\mathrm{aux}}
  + 0.1\,\mathcal{L}_{\mathrm{agr}},
  \label{eq:obj}
\end{multline}
where $\mathcal{L}_{\mathrm{margin}}$ is a bipolar cosine
margin loss on L2-normalised unimodal projections
(pulling congruent pairs together, pushing dissonant pairs
apart); scalar weights were selected by coarse grid search.

\section{CADD-Journal Dataset}
\label{sec:dataset}

Each audio file is rendered by ElevenLabs Eleven~v3~\cite{elevenlabs2025}
with an explicit emotion tag
(negative: \textit{sad, angry, annoyed, appalled};
positive: \textit{happy, excited, cheerful, calm}),
providing acoustic control unavailable in naturalistic speech.
The shared-sentence-pool design renders 100~seed sentences
across all acoustic conditions, so text content is
uninformative for label prediction---any F1 advantage is
genuinely multimodal.
Splits are stratified at the \texttt{sentence\_id} level;
for each dissonance class, 2 emotion tags are randomly drawn
per sentence--voice pair from a pool of 4, yielding
$100 \times 3 \times 2 = 600$ samples per class across
3 voices (Jarnathan, Juniper, Eve); see Table~\ref{tab:dataset}.

\begin{table}[htbp]
\centering\small
\caption{CADD-Journal statistics.}
\label{tab:dataset}
\begin{tabular}{lrrrr}
\toprule
Class & Train & Val & Test & Total \\
\midrule
Masking   & 420 & 90 & 90 & 600 \\
Coping    & 420 & 90 & 90 & 600 \\
Congruent & 420 & 90 & 90 & 600 \\
\midrule
\textbf{Total} & \textbf{1260} & \textbf{270}
  & \textbf{270} & \textbf{1800} \\
\bottomrule
\end{tabular}
\end{table}

\section{Experiments}
\label{sec:experiments}

\subsection{Setup}

\textbf{DACM-Base} replaces asymmetric attention with simple
pooled fusion
($[\mathbf{h}_t - \mathbf{h}_a;\, \mathbf{h}_t \odot \mathbf{h}_a]$),
holding all other components constant to isolate the attention
contribution.
\textbf{DACM-noAttn} retains auxiliary heads and the DIM but
removes asymmetric attention, reverting to pooled fusion
($\mathbf{h}_t$, $\mathbf{h}_a$) as the DIM's input; this
isolates the attention contribution while holding auxiliary
heads and DIM presence constant.
\textbf{DACM-noDIM} applies asymmetric cross-modal attention
but replaces the DIM with direct concatenation of
$\mathbf{h}'_t$ and $\mathbf{h}'_a$, isolating the DIM's
contribution on cross-modal features.
All models use frozen encoders, AdamW ($\beta{=}0.9, 0.999$),
lr~$5{\times}10^{-4}$, batch~16, patience~7, gradient
clipping~1.0, and a static 70/15/15 split.
All results are mean~$\pm$~std over seeds~\{42, 123, 456\}.

\subsection{Results on CADD-Journal}
\label{sec:results}

\begin{table}[htbp]
\centering\small
\caption{Ablation on CADD-Journal test set (3 seeds each).
  Per-class F1 at best seed; Macro = mean~$\pm$~std.
  $^\dagger$TextOnly degenerates to predicting only Congruent
  under shared-sentence-pool conditions; stratified random
  guessing yields macro-F1~$=0.333$.
  $^\ddagger$noAttn: asymmetric attention removed; DIM operates
  on raw pooled $\mathbf{h}_t$, $\mathbf{h}_a$.}
\label{tab:ablation}
\begin{tabular}{lcccc}
\toprule
Model & Mask. & Cop. & Cong. & Macro \\
\midrule
TextOnly$^\dagger$    & 0.000 & 0.000 & 0.500 & $0.167_{\pm 0.000}$ \\
AudioOnly             & 0.630 & 0.535 & 0.255 & $0.448_{\pm 0.023}$ \\
DACM-Base             & 0.628 & 0.562 & 0.183 & $0.441_{\pm 0.014}$ \\
DACM-noAttn$^\ddagger$& 0.642 & 0.578 & 0.254 & $0.469_{\pm 0.021}$ \\
\midrule
DACM-noDIM            & 0.794 & 0.719 & 0.548 & $0.678_{\pm 0.007}$ \\
\textbf{\DACM (ours)}
  & \textbf{0.839} & \textbf{0.845} & \textbf{0.522}
  & $\mathbf{0.711}_{\pm 0.025}$ \\
\bottomrule
\end{tabular}
\end{table}

TextOnly degenerates to predicting a single class (Congruent),
yielding macro-F1~$=0.167$. Note that stratified random guessing
achieves macro-F1~$=0.333$ under our balanced 3-class split;
collapsing to one class is strictly worse because two of the three
class-wise F1 scores are forced to zero. This confirms that text
content alone carries no class-discriminative signal under the
shared-sentence-pool design, so any improvement above this floor
must exploit acoustic information.
DACM-Base ($0.441$) falls below AudioOnly ($0.448$), confirming
that simple pooled fusion actively suppresses the audio
signal---consistent with lexical dominance under
shared-sentence-pool conditions~\cite{chen2026}.

The four-step attribution ladder isolates each component's
contribution.
Adding auxiliary heads and the DIM on top of pooled fusion
(DACM-noAttn, $0.469$) yields only $\Delta{=}{+0.028}$ over
DACM-Base: the DIM is nearly inert when operating on
uncontextualised unimodal features.
Replacing pooled fusion with asymmetric cross-modal attention
--- which provides semantically contextualised features for
the DIM to operate on --- accounts for the dominant gain
($\Delta{=}{+0.242}$, $0.469 \to 0.711$).
The DIM then adds a further $+0.033$ on top of
attention-contextualised features (DACM vs.\ DACM-noDIM),
consistent with it being architecturally motivated for
inference (mismatch score~$S$) rather than as the primary
classification driver.
Gains concentrate on Masking and Coping---the two classes
requiring genuine cross-modal reasoning.
A leave-one-voice-out (LOVO) evaluation yields mean
macro-F1~$0.655$ (Jarnathan: 0.645, Juniper: 0.689, Eve:
0.632), only 0.056 below in-domain, confirming
voice-independent dissonance signatures.

\subsection{Domain Gap: From Controlled TTS to Real Speech}
\label{sec:domaingap}

\begin{table}[htbp]
\centering\small
\caption{Zero-shot cross-corpus evaluation of best \DACM
  checkpoint (in-domain F1~$0.736$).
  $^\dagger$Silver labels via SiEBERT + emotion-to-valence mapping.
  $^\ddagger$2-class (Masking/Coping); random~$0.500$.
  $^\S$3-class; random~$0.333$.
  CH-SIMS uses dataset gold CADD-mapped labels.}
\label{tab:naturalistic}
\begin{tabular}{lllcr}
\toprule
Dataset & Lang. & Size & Label & F1 \\
\midrule
CADD-Journal & EN & 1800  & gold             & 0.736 \\
\midrule
CMU-MOSEI    & EN & 572   & silver$^\dagger$ & 0.512$^\ddagger$ \\
IEMOCAP      & EN & 5674  & silver$^\dagger$ & 0.362$^\S$ \\
CH-SIMS~\cite{yu2020chsims}
             & ZH & 4404  & gold             & 0.335$^\S$ \\
\midrule
Random (2-cls / 3-cls) & & & & 0.500 / 0.333 \\
\bottomrule
\end{tabular}
\end{table}

Table~\ref{tab:naturalistic} is the paper's most practically
significant finding: a model that achieves $0.711$ macro-F1
under controlled conditions collapses to near random on all
three naturalistic corpora (CMU-MOSEI 0.512 over a 0.500
baseline; IEMOCAP 0.362; CH-SIMS 0.335).
This reveals a \emph{severe and previously unquantified}
domain gap between TTS-synthesised dissonance and naturalistic
human speech.
The gap is not merely a model limitation---it reflects the
absence of affective corpora targeting self-reflective
monologues with concealed emotion.
Closing it requires corpora built on two concrete constraints:
(1)~independent dual-channel annotation separating text
sentiment from acoustic valence, and (2)~prospective
user-consented labelling, as concealed states cannot be
annotated reliably post-hoc.

\section{ReflectJournal}
\label{sec:system}

\begin{figure}[htbp]
  \centering
  \includegraphics[width=\columnwidth]{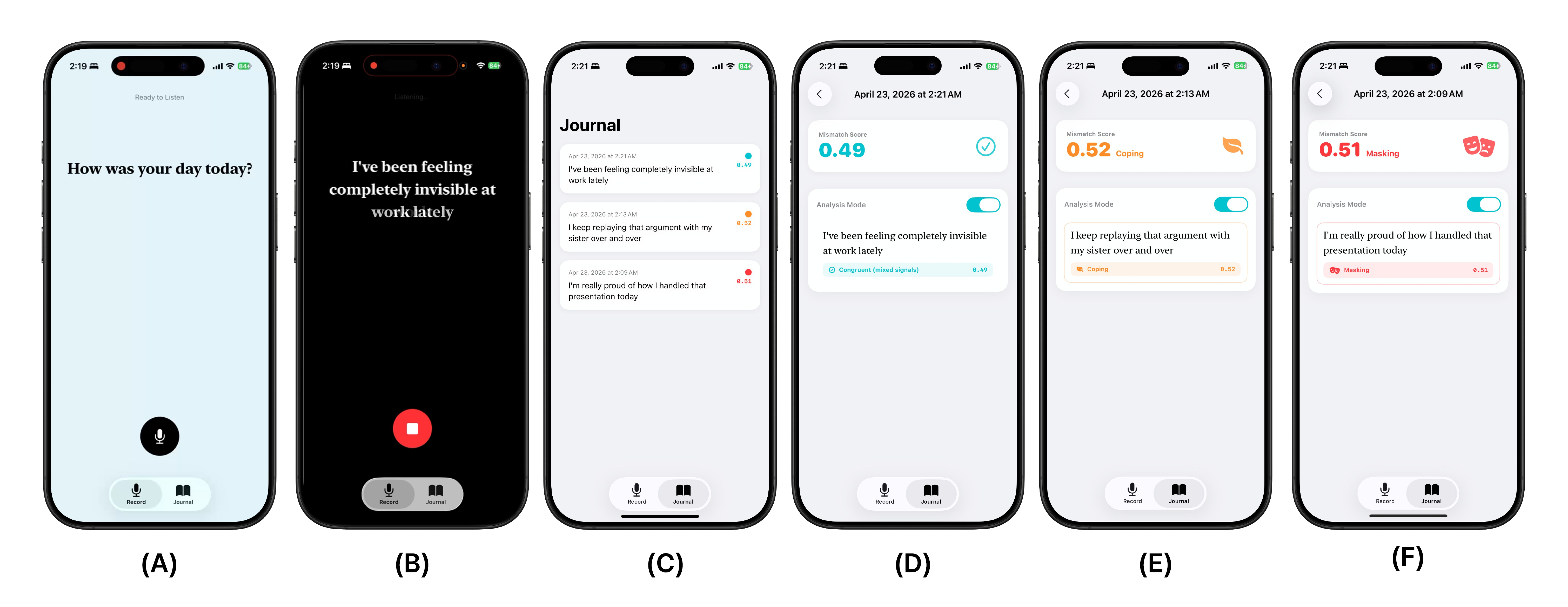}
  \caption{ReflectJournal: (A/B)~recording interface,
    (C)~journal list with mismatch scores,
    (D/E/F)~detail view for congruent, coping, and masking
    states respectively. ``Congruent'' denotes text--audio
    alignment per Table~\ref{tab:taxonomy}.}
  \Description{Six iPhone screenshots arranged left to right.
  Screens A and B show a dark voice-recording interface.
  Screen C shows a journal list with colour-coded mismatch scores.
  Screens D, E, and F each show a single entry detail view
  with a large numeric mismatch score and a coloured class label
  (teal for congruent, orange for coping, red for masking).}
  \label{fig:reflectjournal}
\end{figure}

ReflectJournal is an iOS design probe that operationalizes the
\cadd framework (Figure~\ref{fig:reflectjournal}).
The recording interface (A/B) withholds metrics during capture
to minimise the observer effect; detailed analysis is available
on-demand via a toggle in the journal list (C).
The detail view (D/E/F) surfaces the mismatch score~$S$ and
class-specific reflective prompts when $S > 0.05$,
inviting self-inquiry based on detected cross-modal divergence.
Beyond its journaling function, ReflectJournal serves as a
deployment platform for future in-the-wild corpus collection
with user-consented affective labels, providing a direct pathway
to bridge the TTS-to-real-speech gap identified in
\S\ref{sec:domaingap}.

\section{Discussion and Conclusion}

The central finding of this work is the domain gap: a model
achieving $0.711$ macro-F1 under controlled TTS conditions
collapses to near-random on all naturalistic corpora tested.
This is the expected and scientifically informative outcome for
a first controlled study---it establishes that the CADD
taxonomy is learnable in principle while precisely quantifying
what is missing from current affective corpora.
The Gross-model labels remain interpretive anchors, not
psychological claims; real-speech validation is essential future
work.
The LOVO evaluation (mean F1~$0.655$) provides limited but
positive evidence for voice-independent generalisation across
the three available synthetic voices.

We introduced \cadd to formalize the authenticity gap in
digital journaling, contributed CADD-Journal as a controlled
proxy corpus, and proposed DACM as a baseline architecture.
Most importantly, the cross-corpus evaluation outlines a
concrete research agenda: future work should construct
naturalistic corpora targeting self-reflective monologues with
independent dual-channel annotation and prospective
user-consented labelling.
By providing the task formulation, proxy dataset, baseline
model, and this methodological roadmap, we aim to facilitate
research into genuine cross-modal emotion sensing in the wild.

\section{Limitations}
\label{sec:limitations}

We make our limitations explicit so that subsequent work can
target them directly.

\textbf{Construct validity of TTS-rendered dissonance.}
CADD-Journal isolates acoustic--lexical valence mismatch but
does not instantiate genuine emotion regulation. ElevenLabs
emotion tags impose a controlled mapping between intended
acoustic valence and prosodic realisation; this is a necessary
proxy for studying the signal in a controlled setting, but the
resulting acoustic patterns may differ in distribution from
suppression or reappraisal in human speakers. The Gross-model
labels in Table~\ref{tab:taxonomy} are interpretive anchors,
not psychological claims.

\textbf{Silver labels in cross-corpus evaluation.}
For CMU-MOSEI and IEMOCAP we assigned dissonance labels via
SiEBERT (text sentiment) combined with an emotion-to-valence
mapping (acoustic valence). The disagreement between these two
estimators is itself noisy: a portion of the cross-corpus
performance drop in Table~\ref{tab:naturalistic} is therefore
attributable to label noise rather than to model failure.
Disentangling the two requires gold dissonance labels, which
do not yet exist for these corpora.

\textbf{Voice and language diversity.}
CADD-Journal uses three synthetic English voices. The LOVO
result (mean F1~$0.655$) provides limited evidence for
voice-independent generalisation but cannot speak to broader
demographic variation (age, accent, gender presentation) or to
non-English prosodic systems. The CH-SIMS result reflects both
domain and language transfer and so cannot be uniquely
attributed to either.

\textbf{No user study of ReflectJournal.}
ReflectJournal is presented as a design probe and deployment
platform; we have not yet evaluated whether the mismatch score
$S$, the $S>0.05$ surfacing threshold, or the class-specific
prompts produce useful self-reflection or, conversely, induce
distress in users. The threshold itself was chosen heuristically
and has not been calibrated against human judgement.

\textbf{Seed count and statistical power.}
All ablation results are reported over three seeds
(\{42, 123, 456\}); the resulting standard deviations
yield wide confidence intervals, and small differences such
as DACM~vs.~DACM-noDIM ($+0.033$) should be interpreted as
suggestive rather than firmly established.

\section{Safe and Responsible Innovation Statement}

ReflectJournal infers emotional states from speech, carrying
inherent risks of misinterpretation and user distress if
detections are treated as diagnostic verdicts.
The system is designed exclusively for voluntary self-reflection
in non-clinical populations and must not substitute for
professional mental health support.
CADD-Journal is constructed from synthetically generated audio;
no personally identifiable data is collected or processed.
Dissonance detections are framed explicitly as reflective
starting points, not clinical diagnoses.
Future deployments should prioritise privacy-preserving
on-device inference, establish clinical guidelines
distinguishing therapeutic benefit from potential harm, and
evaluate cross-cultural robustness, as prosodic emotional
norms vary substantially across languages and cultures.

\bibliographystyle{ACM-Reference-Format}
\bibliography{references}

\end{document}